\documentclass[12pt,preprint]{aastex}

\begin{document}

\title{TeV Gamma-Ray Survey of the Northern Hemisphere Sky Using the Milagro Observatory}

\author{R.~Atkins,\altaffilmark{1,10} W.~Benbow,\altaffilmark{2,11} D.~Berley,\altaffilmark{3} E.~Blaufuss,\altaffilmark{3} 
J.~Bussons,\altaffilmark{3,12} D.~G.~Coyne,\altaffilmark{2} 
T.~DeYoung,\altaffilmark{2,3} 
B.~L.~Dingus,\altaffilmark{4} D.~E.~Dorfan,\altaffilmark{2} R.~W.~Ellsworth,\altaffilmark{5} 
L.~Fleysher,\altaffilmark{6} R.~Fleysher,\altaffilmark{6} G.~Gisler,\altaffilmark{4} M.~M.~Gonzalez,\altaffilmark{1} 
J.~A.~Goodman,\altaffilmark{3} T.~J.~Haines,\altaffilmark{4} E.~Hays,\altaffilmark{3} C.~M.~Hoffman,\altaffilmark{4} 
L.~A.~Kelley,\altaffilmark{2} 
C.~P.~Lansdell,\altaffilmark{3}
J.~T.~Linnemann,\altaffilmark{7}
J.~E.~McEnery,\altaffilmark{1,13} R.~S.~Miller,\altaffilmark{8,14}
A.~I.~Mincer,\altaffilmark{6} M.~F.~Morales,\altaffilmark{2,15} P.~Nemethy,\altaffilmark{6} D.~Noyes,\altaffilmark{3} 
J.~M.~Ryan,\altaffilmark{8} F.~W.~Samuelson,\altaffilmark{4} 
A.~Shoup,\altaffilmark{9} G.~Sinnis,\altaffilmark{4} A.~J.~Smith,\altaffilmark{3} G.~W.~Sullivan,\altaffilmark{3} 
D.~A.~Williams,\altaffilmark{2} S.~Westerhoff,\altaffilmark{2,16} 
M.~E.~Wilson,\altaffilmark{1} X.W.~Xu\altaffilmark{4} and G.~B.~Yodh\altaffilmark{9}} 

\altaffiltext{1}{University of Wisconsin, Madison, WI 53706}
\altaffiltext{2}{University of California, Santa Cruz, CA 95064}
\altaffiltext{3}{University of Maryland, College Park, MD 20742}
\altaffiltext{4}{Los Alamos National Laboratory, Los Alamos, NM 87545}
\altaffiltext{5}{George Mason University, Fairfax, VA 22030}
\altaffiltext{6}{New York University, New York, NY 10003}
\altaffiltext{7}{Michigan State University, East Lansing, MI 48824}
\altaffiltext{8}{University of New Hampshire, Durham, NH 03824} 
\altaffiltext{9}{University of California, Irvine, CA 92717}
\altaffiltext{10}{Now at University of Utah, Salt Lake City, UT 84112}
\altaffiltext{11}{Now at Max Planck Institute, Heidelberg, Germany}
\altaffiltext{12}{Now at Universite de Montpellier II, Montpellier, France}
\altaffiltext{13}{Now at NASA Goddard Space Flight Center, Greenbelt, MD 20771}
\altaffiltext{14}{Now at University of Alabama in Hunstville, Huntsville, AL 35899}
\altaffiltext{15}{Now at Massachusetts Institute of Technology, Cambridge, MA 02139}
\altaffiltext{16}{Now at Columbia University, New York, NY 10027}

\begin{abstract}
Milagro is a water Cherenkov extensive air shower array that continuously monitors
the entire overhead sky in the TeV energy band.  The results from an 
analysis of $\sim$3 years of data (December 2000 through November 2003) are presented.  
The data has been searched for steady point
sources of TeV gamma rays between declinations of 1.1 degrees and 80 degrees.  
Two sources are detected, the Crab Nebula and the active galaxy Mrk 421.  
For the remainder of the Northern hemisphere we set 95\% C.L. upper limits between 275 and 600 mCrab
(4.8-10.5$\times 10^{-12}$ cm$^{-2}$ s$^{-1}$) above 1 TeV for 
source declinations between 5 degrees and 70 degrees. Since the sensitivity of
Milagro depends upon the spectrum of the source at the top of the atmosphere,
the dependence of the limits on the spectrum of a candidate source
is presented.   Because high-energy gamma rays from extragalactic sources
are absorbed by interactions with the extragalactic
background light the dependence of the flux limits on the redshift
of a candidate source are given. The upper limits presented here are
over an order of magnitude more stringent than previously published limits from TeV gamma-ray all-sky surveys.  
\end{abstract}



\section{Introduction}
\label{sec:Introduction}

Sources of very-high-energy (VHE, $>$100 GeV) gamma rays are observed to be non-thermal in nature and are 
typically the sites of
particle acceleration.  This acceleration is thought to occur in astrophysical
shocks such as those believed to exist in plerions \citep{deJager1992}, supernova remnants \citep{Volk2000}, 
active galactic nuclei \citep{Blandford1978}, 
and galaxy clusters \citep{Loeb2000} (among other
sources).  These shocks may accelerate protons or electrons, both of which lead to the emission
of gamma rays.  Since gamma rays are unaffected by the magnetic fields that
pervade the Galaxy and the Universe, they can be used to pinpoint the sites of particle acceleration.
In addition to these ``classical'' astrophysical sources of VHE gamma rays, other more 
exotic objects such as primordial black holes, topological defects, and the decay of relic particles from the
big bang may also emit VHE gamma rays.  Perhaps of most interest is the possible existence of
a new type of source that has yet to be postulated.  A comprehensive survey of the sky sensitive to emission
at all time scales is necessary to detect the many possible sources.  
The analysis presented in this paper 
is part of an ongoing effort by the Milagro collaboration to search the entire
northern hemisphere for such objects.  The search for short bursts of TeV gamma rays has been
addressed in a previous paper \citep{Atkins2004}.  The analysis presented here deals specifically with steady 
point sources of TeV gamma rays.  

It has been over a decade since the discovery of the first source of VHE gamma rays, the Crab
Nebula \citep{Weekes89}.  Since that time there have been seven other confirmed sources of TeV gamma rays \citep{Horan2003}, 
six of which lie in the northern hemisphere.  With the exception of the Crab Nebula and the supernova
remnant PKS 1706-44, these objects
are all active galactic nuclei of the blazar class \citep{Horan2003}.   All of these objects have been discovered by atmospheric
Cherenkov telescopes searching for counterparts to sources discovered at lower energies.
In contrast, the EGRET instrument detected over 270 objects emitting high-energy gamma rays above 100 MeV \citep{Hartman99}.
One hundred and seventy of these objects are not identified at other wavelengths.  The VHE regime is a natural
energy band to search for counterparts to these objects.  The small field-of-view and low duty factor
of the atmospheric Cherenkov telescopes (ACTs) make comprehensive sky surveys difficult to perform.
As a result, very few comprehensive surveys of the VHE sky have been performed to date.  
The first VHE survey was performed by a non-imaging ACT
\citep{Helmken79, Weekes79} in 1979.  The Milagrito instrument (a prototype to Milagro, with no background rejection
capability and a higher energy threshold) also performed a survey of the northern hemisphere \citep{Atkins2001} 
and set limits of $\le$3 Crab from any point source in the northern hemisphere.  More recently
flux limits of 4-9 times that of the Crab Nebula (above 15 TeV) have been published by the AIROBICC 
collaboration \citep{Aharonian2002}.  (1 Crab is equivalent
to an integral flux above 1 TeV of $F(>$1 TeV)$=1.75 \times$10$^{-11}$ cm$^{-2}$ s$^{-1}$.) 
The limits presented here are over an order of magnitude more stringent than 
these previous surveys.  

Since many of the confirmed sources of VHE gamma rays are extragalactic the limits
must account for the absorption of TeV 
gamma rays by interactions with the extragalactic background light (EBL) \citep{Primack2000, Stecker2002, Kneiske2002}.  
The EBL is comprised of 
visible radiation emitted by stars and infrared radiation emitted by dust due to reprocessed starlight. 
Since direct measurements
of the intensity and spectrum of the EBL are problematic due to the foreground light
from our galaxy, a model is 
used to determine the effect of the EBL on the observed spectrum at earth from a distant source.  

Before employing the results presented here the limitations
of this survey need to be understood.  First, the limits presented only apply to point sources, 
not to extended objects, such as the galactic plane.  Second, they only apply to the average 
VHE emission during the time period over which the data was obtained.  

\section{The Milagro Observatory}
\label{sec:Milagro}
The Milagro gamma-ray observatory has 723 photomultiplier
tubes (PMTs) submerged in a 24-million-liter water reservoir.  The detector is
located at the Fenton Hill site of Los Alamos National Laboratory, about 35
miles west of Los Alamos, NM, at an altitude of 2630 m asl (750 g/cm$^2$).  The
reservoir measures 80m $\times$ 60m $\times$ 8m (depth) and is covered by a light-tight
barrier.  The PMTs are arranged in two layers,
each on a 2.8m x 2.8m grid.  The top layer of 450 PMTs (under 1.4 meters of
water) is used primarily to reconstruct the direction of the air shower.  By
measuring the relative arrival time of the air shower across the array the
direction of the primary cosmic ray can be reconstructed with an accuracy of
roughly $0.75^\circ$.  The bottom layer of 273 PMTs (under 6 meters of water) is used
primarily to discriminate between gamma-ray-initiated air showers and hadronic
air showers.  

The discrimination of the cosmic ray background is described in detail in \citep{Atkins2003}.
The background rejection uses a parameter known as compactness which is equal to the number of PMTs
in the bottom layer with more than 2 photoelectrons (PEs) divided by the number of PEs in the PMT with
the largest number of PEs (in the bottom layer).
A requirement that the compactness be greater than 2.5 retains 51\% of the
gamma rays while removing 91.5\% of the cosmic ray background (for events with more than 50 PMTs in the
trigger), resulting in an improvement in the gamma-ray
flux sensitivity of 1.7.  Compactness greater than 2.5 was required for this analysis.

Milagro began data taking in 1999.  The data set presented here
begins on 15 December 2000 and ends on 25 November 2003.  Prior to 25 January 2002 the trigger 
was a simple multiplicity trigger requiring 60 PMTs to record a pulse greater than
1/6 of a PE within 180 ns.  After this date a risetime criterion was imposed on the
cumulative timing distribution of struck PMTs.
If a PMT signal is greater than 1/6 of a PE a trigger signal with an amplitude of 6.25 mV and width of 180 ns is generated.
The signals from the 450 PMTs in the top layer are summed and sent to a VME trigger card.  
An 80 MHz flash analog-to-digital converter (FADC) digitizes the trigger signal.  
For each event that exceeds a ``pre-trigger'' condition the data from the FADC
are stored in a FIFO.  The risetime of the trigger signal is defined as the time taken for the trigger signal 
to go from 12.5\% of its peak value to 88.5\% of its peak value.
For events with more than 75 PMTs no risetime requirement was imposed.  Events with more than 52 PMTs 
were required to have a risetime less than 87.5 ns and events with greater than 26 PMTs were 
required to have a risetime less than 50 ns.  The requirement on the risetime removes event triggers due to single muons
at low multiplicity.  The lower multiplicity requirement increases the effective area to 
low-energy events, which improves the sensitivity to gamma-ray bursts.  
The risetime requirement has a minimal impact on the sensitivity to Crab-like sources.  

The trigger rate during this time varied from 1500 Hz to 1800 Hz.  A total of 1009 days of data were acquired
during this period for a detector ``on-time'' of 92\%.  The data are calibrated and reconstructed (to
give the core position, direction, and information used in rejecting the background of the incoming particle)
in real time.  Except for selected regions of the sky, only the reconstructed
information is saved to disk.  This analysis utilizes the reconstructed
data set.  The start date of the analysis is determined by the date that the compactness parameter
and an improved shower-core fitter were included in the reconstructed data.

\section{Analysis Strategy}
\label{sec:Analysis}

Two maps of the sky are constructed: a signal map (comprised of the actual numbers of
events coming from each bin in the sky) and a background map (comprised of an estimate
of the cosmic-ray background from each bin in the sky).  The maps are binned in $0.1\times 0.1$ degree bins.
To estimate the background a technique called ``direct integration" \citep{Atkins2003} is used.  
The method makes use of the fact that the earth rotates and that the
detector response is solely a function of local coordinates and time, and that the cosmic rays constitute an
isotropic background.
The underlying assumption of the method
is that the shape of the detector response (in local coordinates) does not vary over the period during which the
background is accumulated, which is two hours in this analysis.  This method naturally accounts for rate variations
in the detector and makes a high statistics measurement of the background (roughly 12 times as much
background as signal is accumulated for each point in the sky).

A bin of size 2.1 degrees in
declination ($\delta$) by $2.1/\cos(\delta)$ degrees in right ascension is used
to search the skymaps for evidence of a source of TeV gamma rays .  This choice of bin size is
based on the angular resolution of the detector as measured by the shadowing by the Moon of cosmic rays 
and Monte Carlo simulations of gamma-ray showers.  The $0.1\times0.1$
bins in the maps are summed to form these larger bins.  
Large bins are formed with centers at the center of each $0.1\times 0.1$
degree bin by summing the contents of all the small bins that intersect the large bin.  
The summed data from the signal map and background map are then compared.  The declination range of the search
is between 1.1 degrees and 80 degrees. The lower limit is determined by the center of the first large bin with
a low edge of zero degrees declination.  The upper limit is determined by the rotation of the sky, above
80 degrees the hour angle interval used to generate the background is only 2.5 times the size of the signal region.
To calculate the significance of each excess or deficit the prescription of Li and Ma (eq. 17) \citep{LiMa} is used.

\section{Results of Survey}
\label{sec:Results}

Figure 1a shows the distribution of excesses (in standard deviations) for all bins in the
map comprised of the entire 3 years of data.  There is a statistically significant surplus of points
in the sky with greater than 4 standard deviation excess.  The bulk of this surplus can be
attributed to two known sources of TeV gamma rays: the Crab Nebula and the active galaxy Mrk 421.
Figure 1b shows the distribution of excesses with 2-degree regions around the Crab Nebula and Mrk 421 removed.
This distribution is consistent with expectations from random background fluctuations.  A Gaussian fit to this
distribution has a 
mean of -2.9$\times$10$^{-3}$ and a sigma of 0.987, consistent with the number of independent entries in the 
histogram.

Figure 2 shows the map of the northern hemisphere in TeV gamma rays for the data set.
The Crab Nebula (R.A.=83.64, Dec=22.01) and the active galaxy Mrk 421 (R.A.=166.11, Dec=38.21, z=0.03) 
are clearly visible in the map.  The significance of the excess at the location of the Crab is
6.3 standard deviations and at the location Mrk 421 the significance of the
excess is 4.4 standard deviations. (After correcting for instrumental dead-time and other known effects the results 
on the Crab Nebula result in a measured
gamma ray rate of $10.0 \pm 1.4$ events/day.  This is slightly different from the result
given in Atkins et al.(2003) of $10.7 \pm 1.6$ events/day due to the fact that
different calibrations where used in the online reconstruction for some time periods.) 
Table 1 gives the location of all regions
with an excess of greater than 4 standard deviations. Only the pixel with the largest significance from
each independent region is listed in the table.  Near the Crab Nebula the pixel with the greatest
significance is at the location of the Crab Nebula, while near Mrk 421 it is 0.5 degrees from the 
position of Mrk 421.  
The number of these regions is consistent with the expected fluctuations in the background given the large number of trials
incurred in examining the entire sky.  Therefore no claim is made that these regions are sources of TeV gamma rays,
though they may be interesting regions for followup observations with the more sensitive ACTs.
The Whipple Observatory has performed a follow-up observation of the region near R.A.=79.9 Dec=26.8 between November 2002
and January 2003 and has reported
an upper limit of 90 mCrab \citep{Falcone2003}, below the sensitivity of this survey.  

Both the energy response and the sensitivity to gamma ray sources
of Milagro are dependent upon the declination of the source.  
Figure 3 shows the median energy of gamma rays that trigger Milagro (determined from Monte Carlo simulation), 
satisfy the compactness criterion, and are reconstructed
within 1.2 degrees of their true direction, averaged over a complete transit (i.e. 24 hours of observation) 
as a function of the declination of the source for several source spectral indices. The requirement that the
direction of the particle be reconstructed within 1.2 degrees of its true direction is imposed to account for the
bin size used in the analysis (a 2.1 degree wide square bin has the same area as a 1.2 degree circular bin).  

Establishing upper limits to the gamma ray flux from any given 
point in the sky is straightforward for galactic (nearby) sources of TeV gamma rays.  
The prescription of Helene \citep{Helene83} is used to calculate 
the confidence limits on the number of signal events from each region of the sky.
Since the response of Milagro is energy dependent, the flux upper limits obtained from these data
are dependent upon the energy spectra of the possible sources of TeV gamma rays.  Figure 4 is a 
2-dimensional map of the sky with the 95\% C.L. upper limits to the flux given at each point.  
These limits are based on the assumption of source spectra proportional to $E^{-2.59}$ (i.e. similar to the 
Crab Nebula \citep{Aharonian2002, Atkins2003} at these energies).  
In order to translate the observed upper limits on the number of excess events from a given location into
an upper limit on the flux of gamma rays, the detector response is normalized using the results from the Crab Nebula.
This procedure accounts for the dead time of the detector, calibration errors, and other systematic effects. 
At declinations near 36 degrees (the latitude of Milagro) the 95\% C.L. upper limits are on average 275 mCrab or
$F$($>$1 TeV)$<$4.8$\times$ 10$^{-12}$ cm$^{-2}$ s$^{-1}$.  As the declination of the source increases or
decreases from this value the source spends less time near zenith (where the atmospheric overburden is least
and the response of Milagro is best) and the flux upper limits increase.  At declinations of 5 degrees or 65 degrees 
the average 95\% C.L. upper limits are of the order 600 mCrab ($F(>$1 TeV) $<$ 1.05$\times$ 10$^{-11}$ cm$^{-2}$ s$^{-1}$).
Figure 5 gives the factor by which these upper limits must be multiplied for potential sources with different
spectral indices. The use of this figure is best
explained with an example.  From Figure 4 obtain an upper limit from a location, R.A.=180., Dec=40 ($F(>$1 TeV)$< 291$ mCrab).
To find the flux upper limit for a source with a differential spectral index of -2.0, find the curve for such a
source in Figure 5 (the solid curve).  At the declination of the source the y-axis value (0.41) 
is the amount by which the flux upper limit
from Figure 4 must be multiplied to give the flux upper limit for this source (119 mCrab).

For an extragalactic source one must also account for the absorption of the gamma rays due to interactions
with the EBL \citep{Primack2000, Stecker2002, Hartman99}.  In the absence of a
reliable measurement of the EBL, a model of the intensity and energy spectrum is used.  Figure
6 shows the effect of the absorption due to the EBL on the upper limits given in Figure 4.
In this figure the ``baseline'' model of Stecker and De Jager \citep{Stecker2002} is used to calculate the
effect of the absorption of energetic gamma rays and the source spectra are assumed to be
proportional to $E^{-2.59}$.  Using a procedure similar to that
described above for Figure 5, one can find the flux upper limit for sources at different redshifts.  
For example, using the same location as above (R.A.=180 Dec=40) but a source at a redshift of 0.03 (with a 
differential spectral index of -2.59), the curve for z=0.03 in Figure 6 
has a value of ~2.4, resulting in a flux upper limit for this location and redshift of 700 mCrab.
This upper limit is the normalization of the power law spectrum {\em of the unabsorbed source} at the top
of the atmosphere.  Assuming a source with an intrinsic spectrum (before absorption by the EBL)  represented by,
\begin{equation}
\frac{dN}{dE} = I_0 {E_{TeV}}^{-\alpha}
\end{equation}
then, for this source $I_0$ would be 2.4 times that of a galactic source with the same spectral index.
(Note that to calculate the absolute luminosity of the source
one must also multiply by the square of the distance to the source.  This has not been accounted for in Figure 6.)
Because the absorption of TeV gamma rays by the EBL distorts the source spectrum before
it reaches the earth one can not use Figure 6 in series with Figure 5.  In general the effect on the sensitivity due to 
different source spectra is smaller for distant sources, since the EBL tends to make the distant sources look more alike
regardless of their spectra.  Figure 7 gives an example of the relative sensitivity to sources at different redshifts with
different spectra.  Figure 7a is similar to Figure 6 but the sources here are assumed to have a differential photon spectrum
proportional to $E^{-2}$ and in Figure 7b a spectrum proportional to $E^{-3}$.  The figure gives the ratio of the flux upper limit
for a source at the given redshift, declination, and spectrum to a local source (z=0.0) with a differential spectral index of
-2.59.  A perhaps surprising feature of this figure is that for
more distant sources a source with a harder intrinsic spectrum is required to have a larger luminosity than a source with a softer
spectrum for Milagro to make a detection.  This is due to the effect of the EBL, where the high energy photons that the source
is required to emit (by the model) are absorbed in transit and do not affect the ability of Milagro to observe the source
but do count as part of the intrinsic source luminosity.

\section{Conclusions}
\label{sec:Conclusions}
A complete survey of the northern hemisphere (declination $1.1^{\circ}$ to $80^{\circ}$) for point sources
of TeV gamma rays has been performed.  These limits apply to the average flux level during the roughly 
3 year period from December 15, 2000 
through November 25, 2003.  The average 95\% C.L. upper limits range from 275 mCrab to 600 mCrab depending upon the 
declination of the source and are over an order of magnitude more restrictive than previous limits.  
A prescription has been given to calculate the corresponding
upper limits for sources with different spectra and for extragalactic sources.  For sources with differential spectral
indices of -2.0 the upper limits are 57\% lower.  For a source at a redshift of 0.03
the flux limits are a factor of 2.4 larger.  While these limits are the best available to date, 
Milagro has recently been completed with the construction of an array of 175 water tanks surrounding 
the central reservoir.  A comparable dataset, with this now complete Milagro detector, 
would improve these limits by a factor $\sim$2.

\acknowledgments

We gratefully acknowledge Scott Delay and Michael Schneider for their dedicated efforts in the 
construction and maintenance of the Milagro experiment.  This work has been supported by the 
National Science Foundation (under grants 
PHY-0070927, 
-0070933, 
-0075326, 
-0096256, 
-0097315, 
-0206656, 
-0302000, 
and
ATM-0002744) 
the US Department of Energy (Office of High-Energy Physics and 
Office of Nuclear Physics), Los Alamos National Laboratory, the University of
California, and the Institute of Geophysics and Planetary Physics.

\clearpage

\begin{figure}
  \figurenum{1}
  \epsscale{1}
\plotone{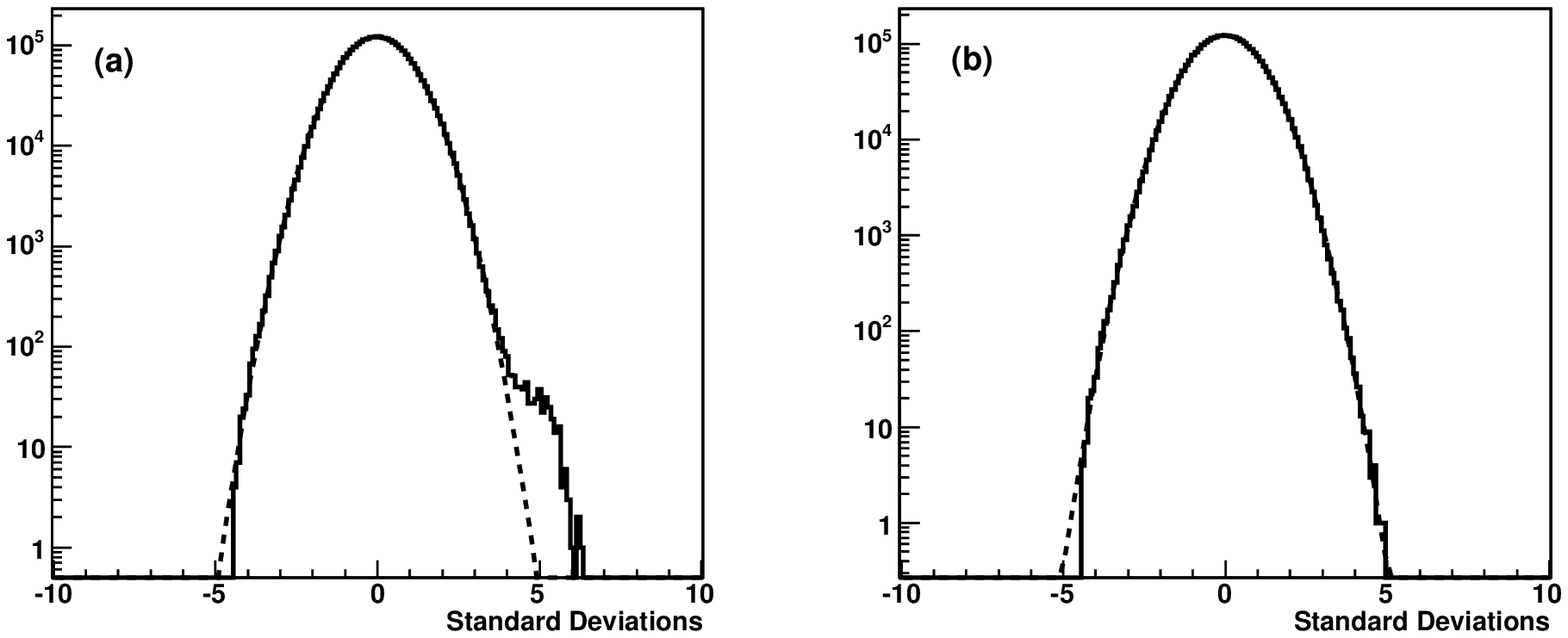}
  \caption{(a) The distribution of significances of the excesses and deficits in the analysis of the
  D.C. skymap of the northern hemisphere.  (b) The same data with 2 degree regions around
the Crab Nebula and Mrk 421 removed.  The dotted curves are the best fit Gaussians to the data.}
\end{figure}
                                                                                                                             
\begin{figure}
  \figurenum{2}
\plotone{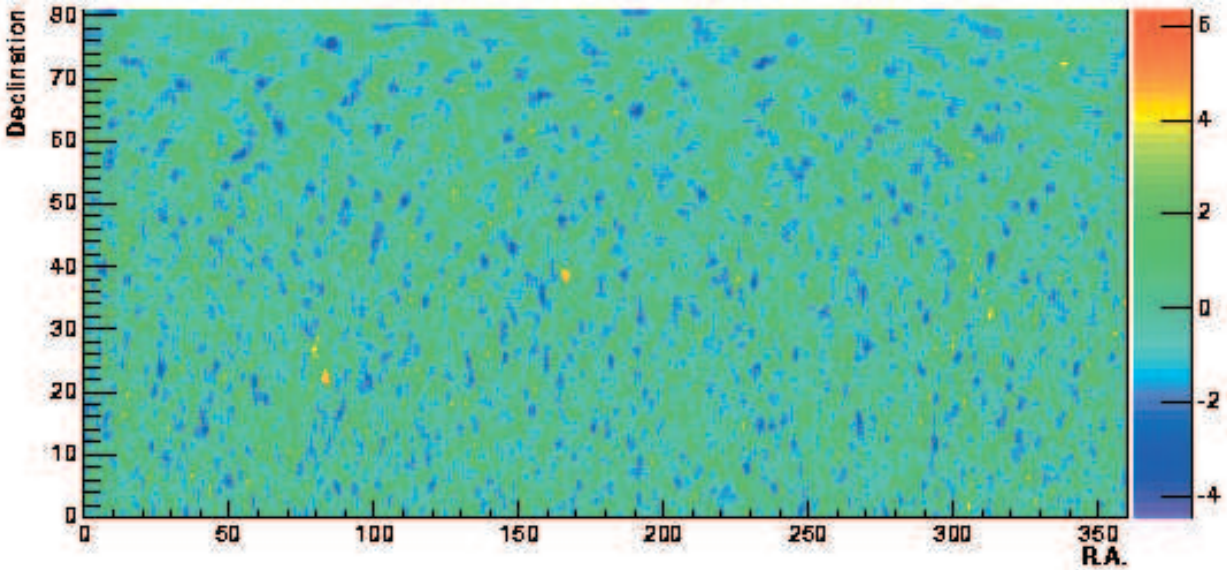}
  \caption{The northern hemisphere as seen in TeV gamma rays.  At each point the excess is summed over
  a 2.1 degree by $2.1/\cos(\delta)$ bin and the significance of the excess in standard deviations is 
  shown by the color scale.}
\end{figure}

\begin{figure}
  \figurenum{3}
\plotone{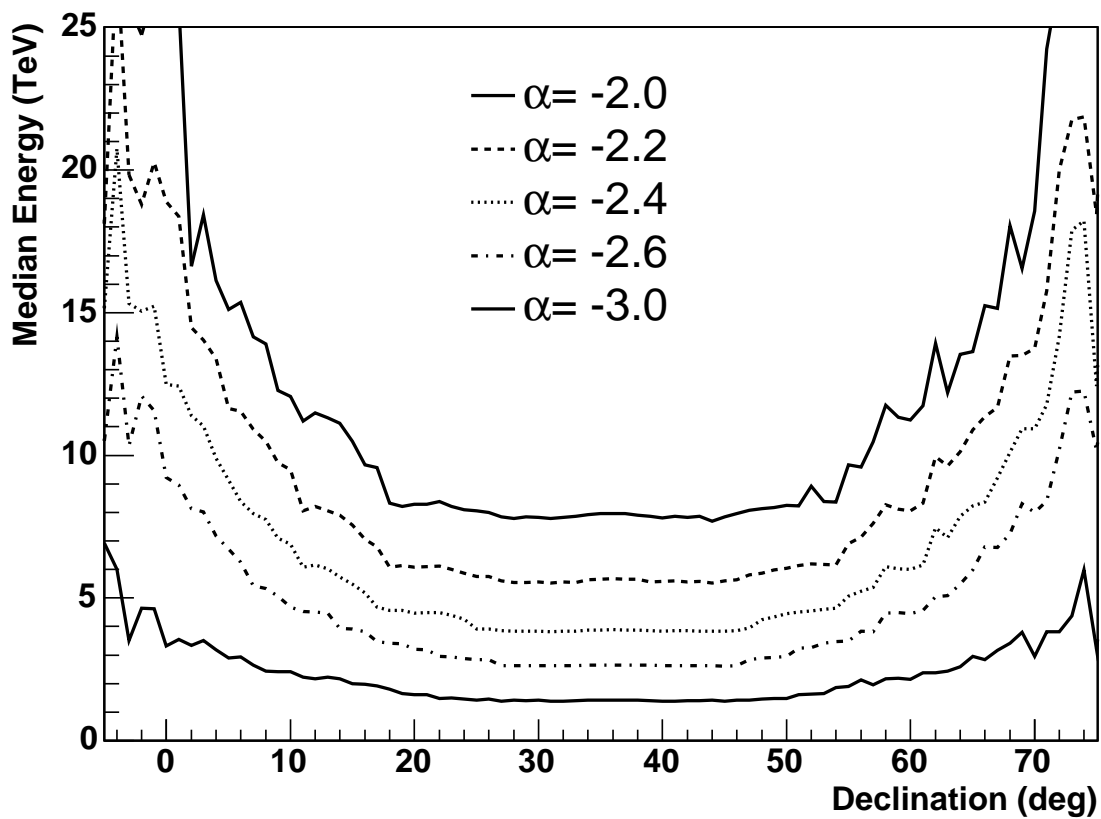}
  \caption{The median energy of gamma-ray events that trigger Milagro, pass the compactness cut, and 
are reconstructed with 1.2 degrees of their true direction as a function of source declination (this is 
equivalent to the 2.1 degree square bin used in the search).  The response
of Milagro is averaged over a complete source transit (i.e. one day's observation)
and the source differential spectral index, $\alpha$, for each curve is given in the legend.}
\end{figure}

\begin{figure}
  \figurenum{4}
\plotone{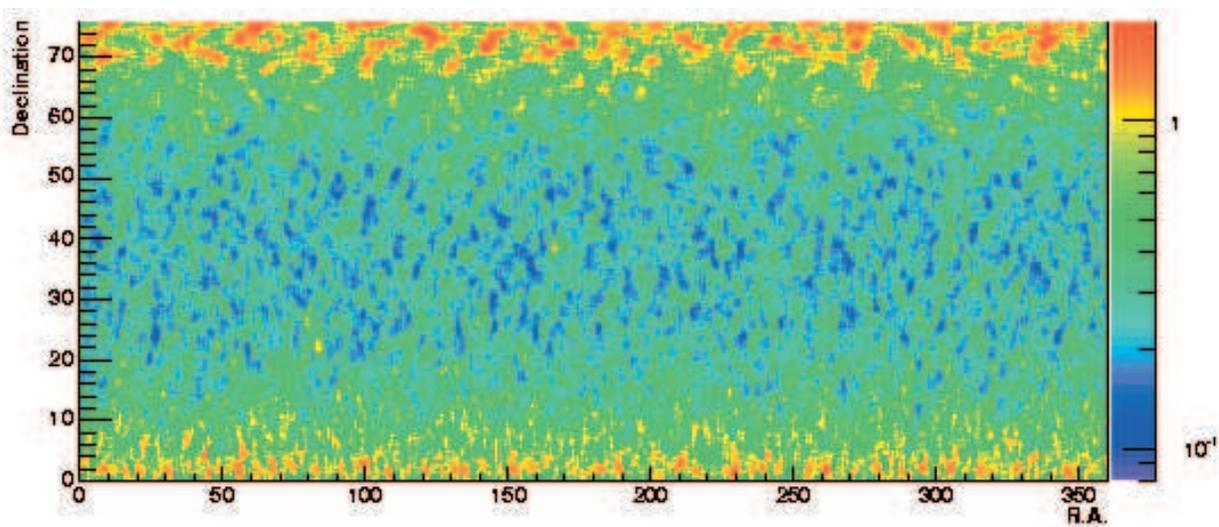}
  \caption{The 95\% C.L. upper limits on the integral flux of gamma rays above 1 TeV (assuming an $E^{-2.59}$
differential photon spectrum) from each point in the northern hemisphere.  The color scale on the right
is in units of the flux from the Crab Nebula.  To enhance the contrast of the figure only declinations below 75 degrees
are shown.}
\end{figure}

\begin{figure}
  \figurenum{5}
\plotone{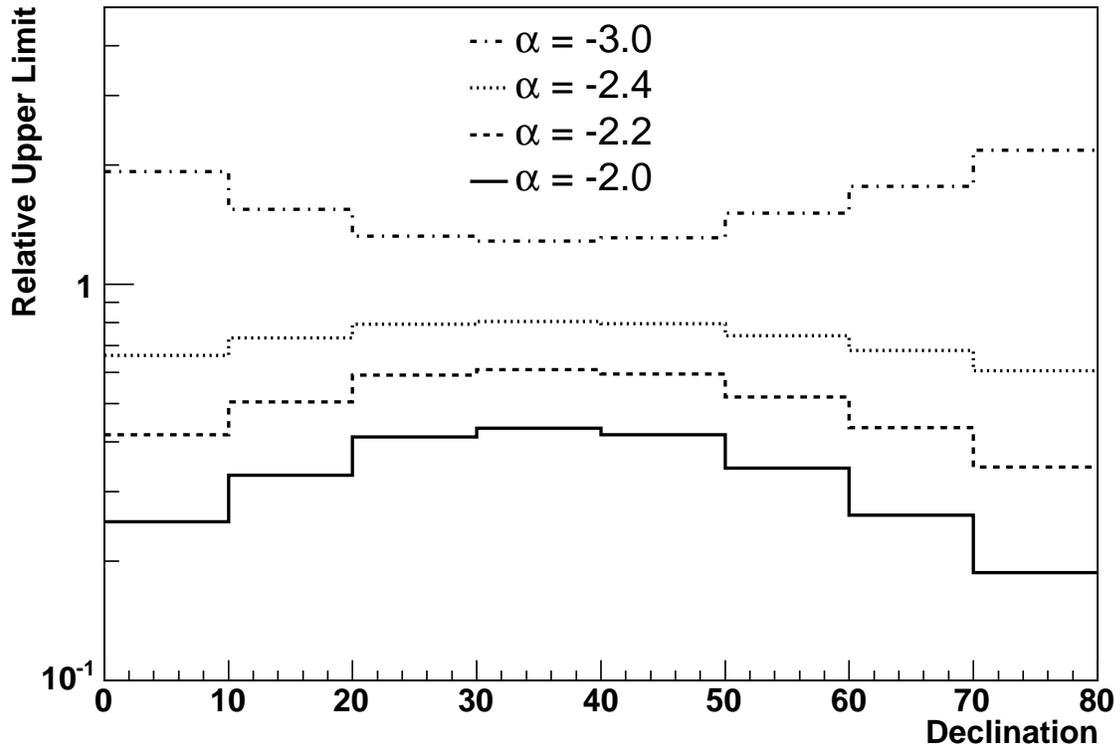}
  \caption{The effect of the differential spectral index on the upper limits shown in Figure 4.   
The y-axis gives the ratio of the flux upper limit for a source with a differential spectral index as
indicated by the curve to a source with a differential spectral index of -2.59.  The use of this figure is described 
in the text.}
\end{figure}

\begin{figure}
  \figurenum{6}
\plotone{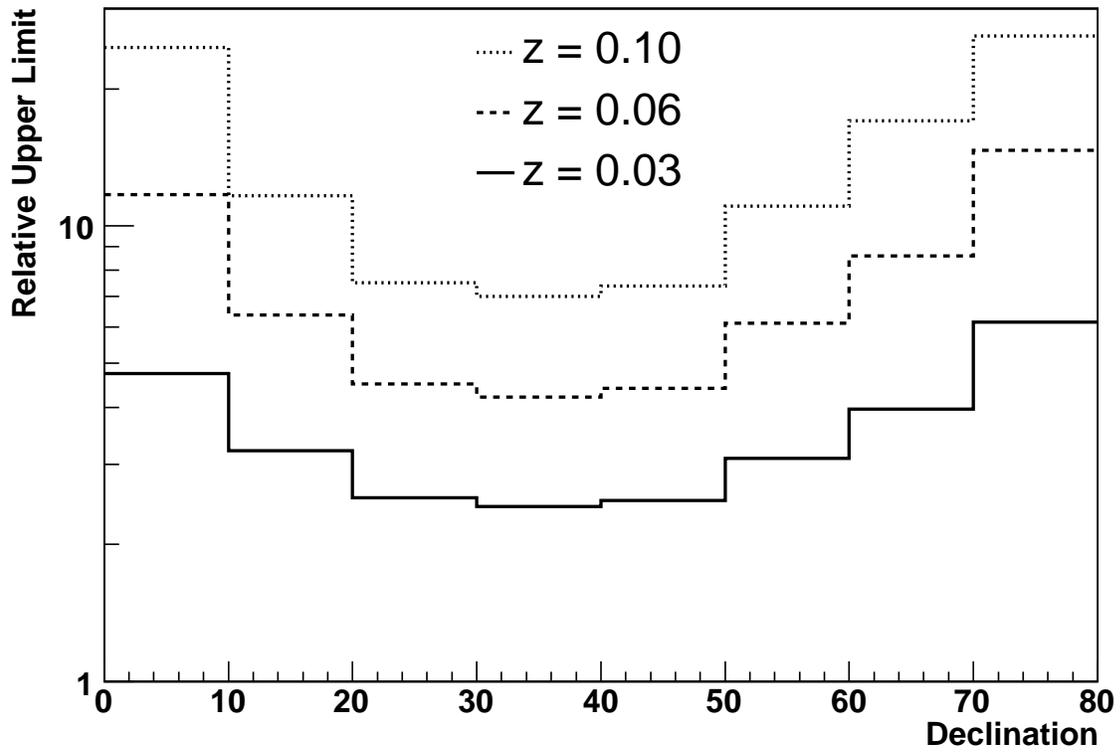}
  \caption{The effect of redshift on the upper limits shown in Figure 4.  These results
assume a source spectrum proportional to $E^{-2.59}$.  The y-axis is the ratio of the flux upper limit for a source
at the indicated redshift to a source at a redshift of zero.}
\end{figure}

\begin{figure}
  \figurenum{7}
\plotone{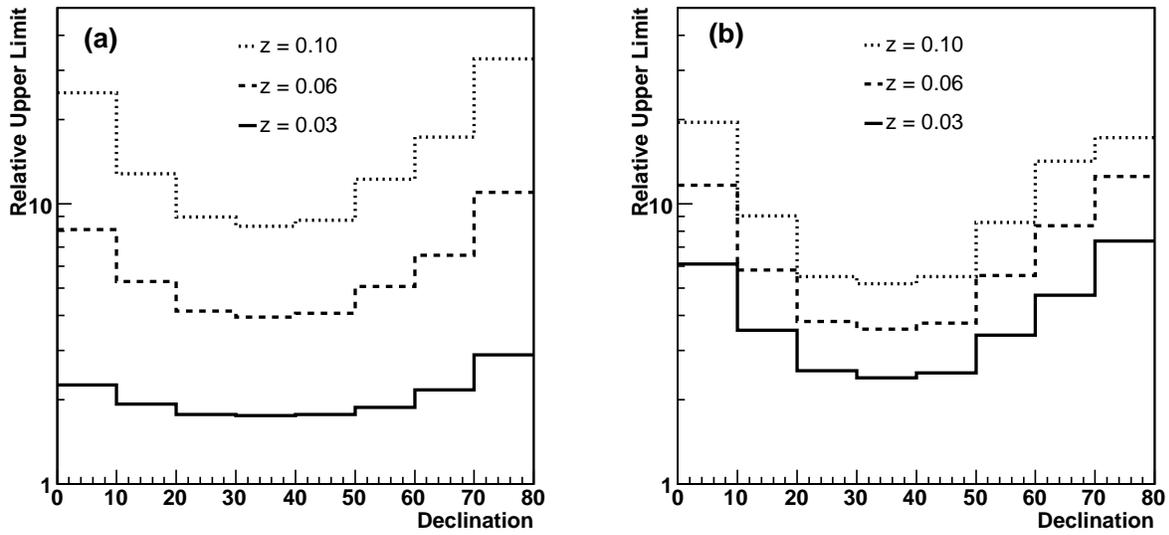}
  \caption{The effect of redshift on the upper limits shown in Figure 4.  Figure 7a
assumes a source spectrum proportional to $E^{-2}$ and 7b a source spectrum proportional to $E^{-3}$.
In both cases the y-axis is the ratio of the flux upper limit for the described source 
(spectral index, redshift, and declination) and a local source with a differential spectral index of -2.59.}
\end{figure}

\clearpage

\begin{deluxetable}{ccccccc}
\tabletypesize{\scriptsize}
\tablecaption{The locations of all regions with greater than a 4$\sigma$ excess.
Only independent regions are entered in the table.  Errors on the location
of any possible source are $\sim$0.5 degrees.  The two brightest points on the map
are due to the two detected sources: the Crab Nebula$^1$ and Mrk 421$^2$.  Upper
limits are not given for these two sources.
The units of right ascension and declination are decimal degrees.
The last column gives the 95\% C.L. upper limit to the flux in units of the Crab
flux.  \label{tbl-1}}
\tablewidth{0pt}
\tablehead{
\colhead{RA} & \colhead{DEC}   & \colhead{ON}   &
\colhead{OFF} & \colhead{Excess} & \colhead{Sigma} & \colhead{UL} 
}
\startdata
0.3       &   34.3 &  3.12308e+06 &    3.11456e+06  &   8623   &  4.7  & 0.84 \\
37.8      &   6.7  &  7.02166e+05 &    6.98667e+05  &   3498   &  4.0  & 1.8 \\
43.6      &   4.8  &  5.85952e+05 &    5.82716e+05  &   3236   &  4.1  & 2.0 \\
49.1      &   22.5 &  2.21431e+06 &    2.20813e+06  &   6175   &  4.0  & 0.87\\
79.9      &   26.8 &  2.57841e+06 &    2.57025e+06  &   8161   &  4.9  & 0.97\\
83.6$^1$  &   22.0 &  2.17188e+06 &    2.16222e+06  &   9665   &  6.3  & NA \\
166.5$^2$ &   38.6 &  3.23552e+06 &    3.22467e+06  &   10850  &  5.8  & NA \\
306.6     &   38.9 &  3.25329e+06 &    3.24531e+06  &   7983   &  4.2  & 0.78\\
313.0     &   32.2 &  3.08380e+06 &    3.07548e+06  &   8320   &  4.5  & 0.85\\
339.1     &   72.5 &  6.63534e+05 &    6.59727e+05  &   3807   &  4.2  & 3.02\\
356.4     &   29.5 &  2.98656e+06 &    2.97910e+06  &   7455   &  4.1  & 0.84\\
\enddata
\end{deluxetable}


\begin{thebibliography}{}

\bibitem[Aharonian et al.(2002)]{Aharonian2002} Aharonian, F., et al., 2002, \aap, 390, 39
\bibitem[Atkins et al.(2001)]{Atkins2001} Atkins, R., et al., 2001, \apj, 558, 477
\bibitem[Atkins et al.(2003)]{Atkins2003} Atkins, R., et al., 2003, \apj, 595, 803
\bibitem[Atkins et al.(2004)]{Atkins2004} Atkins, R., et al., 2004, \apjl, to appear
\bibitem[Blandford and Ostriker(1978)]{Blandford1978} Blandford, R.~D. and Ostriker, J.~P., 1978, \apjl, 221, 29
\bibitem[De Jager and Harding(1992)]{deJager1992} De Jager, O.~C. and Harding, A.~K., 1992, \apj, 396, 161
\bibitem[Falcone et al. (2003)]{Falcone2003} Falcone, A., et al., 2003, in Proc. 28th International Cosmic Ray Conference, ed. T. Kajita, Y. Asaoja, A. Kawachi, Y. Matsubara, and M. Sasaki (Tokyo, Japan), Vol. 5, 2579
\bibitem[Li and Ma(1983)]{LiMa} Li, T.~P. and Ma, Y.~Q., 1983, \apj, 272, 317
\bibitem[Loeb and Waxman(2000)]{Loeb2000}Loeb, A. and Waxman, E., 2000, Nature, 405, 156
\bibitem[Hartman et al.(1999)]{Hartman99} Hartman, R.~C., et al., 1999, \apjs, 123, 79
\bibitem[Helene(1983)]{Helene83} Helene, O., 1983, NIM, 212, 319
\bibitem[Helmken et al.(1979)]{Helmken79} Helmken, H.~F., Horine, E., and Weekes, T.~C., 1979, Proceedings of the 19th ICRC, 1, 120
\bibitem[Horan and Weekes(2003)]{Horan2003} Horan D. and Weekes, T.~C., 2003, astro-ph/030391v1
\bibitem[Kneiske, Mannheim, and Hartmann(2002)]{Kneiske2002} Kneiske, T.M., Mannheim, K., and Hartmann, D.H., 2002, \aap, 386, 1
\bibitem[Primack et al.(2000)]{Primack2000} Primack, J.~R., Somerville, R.~S., Bullock, J.~S., and Devriendt, J.~E.~G., 2000,AIP Conference Proceedings, 558, 463, AIP: F.~A. Aharonian and H.~J. V{\" o}lk
\bibitem[Stecker and De Jager(2002)]{Stecker2002} Stecker, F. and De Jager, O.~C., 2002, \apj, 566, 738
\bibitem[Berezhko and V{\" o}lk(2000)]{Volk2000}Berezhko, E.~G. and V{\" o}lk, H.~J., 2000, APh, 14, 201
\bibitem[Weekes et al.(1979)]{Weekes79} Weekes, T.~C., Helmken, H.~F., and L'Heureux, J., 1979, Proceedings of the 19th ICRC, 1, 126
\bibitem[Weekes et al.(1989)]{Weekes89} Weekes, T.~C., et al., 1989, \apj, 342, 379
\end{thebibliography}
\end{document}